\documentclass[namedreferences]{SolarPhysics}
\usepackage[optionalrh,solaenum]{spr-sola-addons} 

\usepackage[pdftex]{graphicx}
\usepackage{epstopdf}

\usepackage{color}
\usepackage{url}

\newcommand{\apj}{  {\it Astrophys. J.}}
\newcommand{\apjl}{  {\it Astrophys. J. Lett.}}
\newcommand{\apss}{  {\it Astrophys. Space Sci.}}

\newcommand{\jgr}{  {\it J. Geophys. Res.}}

\newcommand{\solphys}{{\it Solar Phys.}}

\newcommand{\cref}[1]{Chapter~\ref{#1}}
\newcommand{\sref}[1]{Section~\ref{#1}}
\newcommand{\eref}[1]{Equation~(\ref{#1})}
\newcommand{\fref}[1]{Figure~\ref{#1}}

\newcommand\pdf{pdf~}
\newcommand\wt{wt~}

\newcommand{\ptp}{\textit{PtP~}}
\newcommand{\ets}{\textit{EtS~}}

\newcommand{\Ha}{H$\alpha$~}
\newcommand{\Xr}{$X$-ray~}
\newcommand{\etal}{{\it et al.}}

\begin{document}
\begin{article}
\begin{opening}

\title{Solar Flare Occurrence Rate and Waiting Time Statistics}

\author{A.~\surname{Gorobets}$^{a}$}
\institute{$^{a}$Sterrekundig Instituut, Utrecht University, Postbus 80000, NL-3508 TA Utrecht, The Netherlands\\
  email: \url{a.y.gorobets@astro-uu.nl}}

\author{M.~\surname{Messerotti}$^{b,c}$}
\institute{$^{b}$ INAF-Astronomical Observatory of Trieste, Loc. Basovizza 302, 34012 Trieste, Italy}
\institute{$^{c}$ Department of Physics, University of Trieste, Via A. Valerio 2, 34127 Trieste, Italy}

\begin{abstract}
We use Renewal Theory for the estimation and interpretation of the flare rate from the \textit{Geostationary Operational Environmental Satellite} (GOES) soft \Xr flare catalogue. It is found, that in addition to the flare rate variability with the solar cycles, a much faster variation occurs. The fast variation on time scales of days and hours down to minute scale appears to be comparable with time intervals between two successive flares (waiting times). The detected fast non-stationarity of the flaring rate is discussed in the framework of the previously published stochastic models of the waiting time dynamics.
\end{abstract}


\end{opening}

\section{INTRODUCTION}\label{intro}

The phenomena related to energy transformation and release on the Sun, possibly are of highest importance for modern solar astrophysics. Solar flaring, as a steady process of energy release, plays a central role and has been drawing the attention of the scientific community for decades.

In this work, we focus on the statistical properties of so-called solar flares \textit{waiting time(s)} (wt), \textit{i.e.} the interval between two flares close in time. The data are provided by the \textit{Geostationary Observational Environmental Satellites} (GOES) in the soft \Xr band. This catalogue is chosen as the longest available record of uninterrupted observations. The \Ha flare  records from the NGDC-NOAA catalogue$\footnote{http://www.ngdc.noaa.gov/stp/solar/solarflares.html}$ are shorter, and used to emphasise the invariance of the reported results.

Flare waiting times statistics has been extensively debated in the literature. \inlinecite{Boffetta} fitted the waiting time probability density function (pdf)\footnote{We use the mathematically strict definition of the \textit{density} as the differential quantity with respect to the \textit{distribution}, which is often used in a misleading way instead.} with a power-law in the range between $6$ and $67$ hours. This estimate was implemented using a single record over $20$ years. \inlinecite{Wheat-II} considered the variation of the mean flaring rate with the solar cycle, and proposed the model of the time-dependent Poisson process (see also \opencite{Moon2001}). In turn, this model gives a power-law-like \pdf only in the limit \cite{Wheat-final}. However, \inlinecite{Lepreti} argued for a local departure of the \wt series from the Poisson process, \textit{i.e.} "memory" had been detected in the data.

We apply an alternative statistical description to the methods that had been used in the articles cited above. This method is equivalently unambiguous in describing the stochastic processes. The flare rate is estimated explicitly from the waiting time pdf. Such an approach has the significant advantage of linking the waiting time of the flare and its rate, which is the instantaneous probability of the flare per unit of time.

The paper is organised as follows. In Section~\ref{math} the rather simple mathematics used in the paper is summarised, considered for the paper's consistency to avoid referring the reader to specific literature. The main results are presented Section~\ref{data_analysis}, and interpreted in Section~\ref{discussion}. The conclusion and final remarks are presented in Section~\ref{conclusion}.

\section{THEORETICAL BACKGROUND}\label{math}
We adopted the terminology of Renewal Theory (\emph{e.g.}~\opencite{Cox}) to set a mathematical framework and for intuitive and easy further interpretation.

We focus on two objects of study: random events and time interval between near-in-time events, \textit{i.e.}, \textit{waiting times}. By analogy with Renewal Theory, the \textit{failure} of some abstract device is attributed as the elementary random \textit{event}, which we assign to the \textit{flare}. Then the \textit{flare waiting time} is associated with a random non-negative variable $X$, called \textit{failure time}\footnote{This time interval is also called the \textit{age} of a device, \textit{i.e.,} its lifetime without any failure.} being the interval between adjacent\footnote{An abstract picture is considered here: a broken device replacement (repair) is thought to be a very instant, in such a way that the time required for the repair is zero. A continuous operation is meant, and provided, for instance, by multiple hardware availability.} \textit{failures}, that is, between two flares.

The random variable $X$ is characterised by a \textit{probability density function}~$f(x)$

\begin{equation}\label{eq:pdf}f(x)= \lim_{\Delta x\rightarrow0} \frac{Prob (x\leq X \leq x+\Delta x)}{\Delta x},\end{equation}
\noindent
with $\int_{0}^{\infty} f(x) dx = 1.$

The probability that a flare \textit{has} occurred (a device has failed) by time $x$ is given by the \textit{cumulative distribution function} $F(x)$:

\begin{equation}\label{eq:cum}F(x)=Prob(X \le x)=\int_{0}^{x} f(u) du.\end{equation}

The probability that a flare \textit{has not} occurred (a device has not failed) up to time $x$ is given by the \textit{survivor function} $\mathcal{F}(x)$:

\begin{eqnarray}
\mathcal{F}(x) &=& Prob(X > x) = \\
 1-F(x) & = &\int_{x}^{\infty} f(u) du.
\end{eqnarray}\label{eq:sur}

The probability of immediate failure of a device (occurrence of a flare event) known to be of age $x$ (no flare during $x$) is given by the \textit{age-specific failure rate} $h(x)$\footnote{This function is also known as the \textit{hazard function}, \textit{hazard rate} or \textit{failure rate}. }. Consider a device known not to have failed at time $x$ and let $h(x)$ be the limit of the ratio to $\Delta x$ of the probability of failure
in time interval~$(x, x +\Delta x]$

\begin{equation}\label{eq:h}h(x)= \lim_{\Delta x\rightarrow 0} \frac{Prob(x<X \leq x+\Delta x|x<X)}{\Delta x}.\end{equation}

The latter permits further transformation according to the definition of the conditional probability for two events $a, b,$

\begin{equation}\label{eq:cond}Prob(a|b)=\frac{Prob(a \mbox{ and } b)}{Prob(b)}.\end{equation}

The event $$"x<X \leq x+\Delta x \mbox{ \textit{and} } x<X"$$ is included in essentially the same event $$"x<X \leq x+\Delta x"\:;$$ then \eref{eq:h} reads 

\begin{eqnarray*}
\lefteqn{h(x)=}\\
&&=\lim_{\Delta x\rightarrow0} \frac{Prob(x<X \leq x+\Delta x)}{\Delta x}\frac{1}{Prob(x<X)}\\
&&=\frac{f(x)}{\mathcal{F}(x)}
\end{eqnarray*}\label{eq:h1}

and the next form is used for the computations in the paper:

\begin{equation}\label{main_eq}h(x)=\frac{f(x)}{1-F(x)}=\frac{f(x)}{1-\int_{0}^{x} f(u) du}.\end{equation}

The stochastic \textit{Poisson process} has the exponential waiting time \pdf \begin{equation}\label{exp}f(x)=\lambda e^{-\lambda x},\end{equation}
where $\lambda=<x>^{-1}$ is the \textit{constant} failure rate
\begin{equation}\label{hexp}h_p(x)= \frac{\lambda e^{-\lambda x}}{\int_{x}^{\infty} \lambda e^{-\lambda u}du}=
\lambda.\end{equation}

Conversely, if the stochastic process has a constant $h$, it is a Poisson process. The constancy of $h_p(x)$ reveals the "no-memory" property
of the Poisson stochastic process.

\section{DATA ANALYSIS}\label{data_analysis}

The GOES flare catalogues are analysed with particular interest in the interval between consecutive flares.
The GOES observations provide the longest record and complementary H$\alpha$ data provided allow one to consider both \Xr and \Ha flares.

The catalogued event is recorded by specifying starting time, peak time and ending time. Thus, the waiting time is not unambiguously defined. We consider two definitions: the \textit{Peak-to-Peak} (PtP) waiting time\textemdash the interval between times of maximum flux rate of two near-in-time flares, and the \textit{End-to-Start} (EtS) waiting time\textemdash the interval between the end time of the predecessor and the starting time of the current event. The latter definition is intuitive, but might be of questionable applicability from the point of view of statistical methods that will be used (see Section~\ref{discussion}). However, it is used in the article to demonstrate the influence of the definition of the \wt on its statistical properties.

Initially, the catalogue should be preprocessed to avoid errors due to data gaps. The neighbourhood events to every catalogue item marked with the data gap flags (notably \texttt{D}, \texttt{E}) were ignored; thus, after filtering we have the set of $52181$ \Xr events (including GOES $B$-class events) dating from September $1975$ until December $2008$. The \Ha events record is shorter, \emph{viz.} $11124$ events and it is used only for numerical comparison in the following.

\begin{figure}
\includegraphics[width=1\textwidth,clip=]{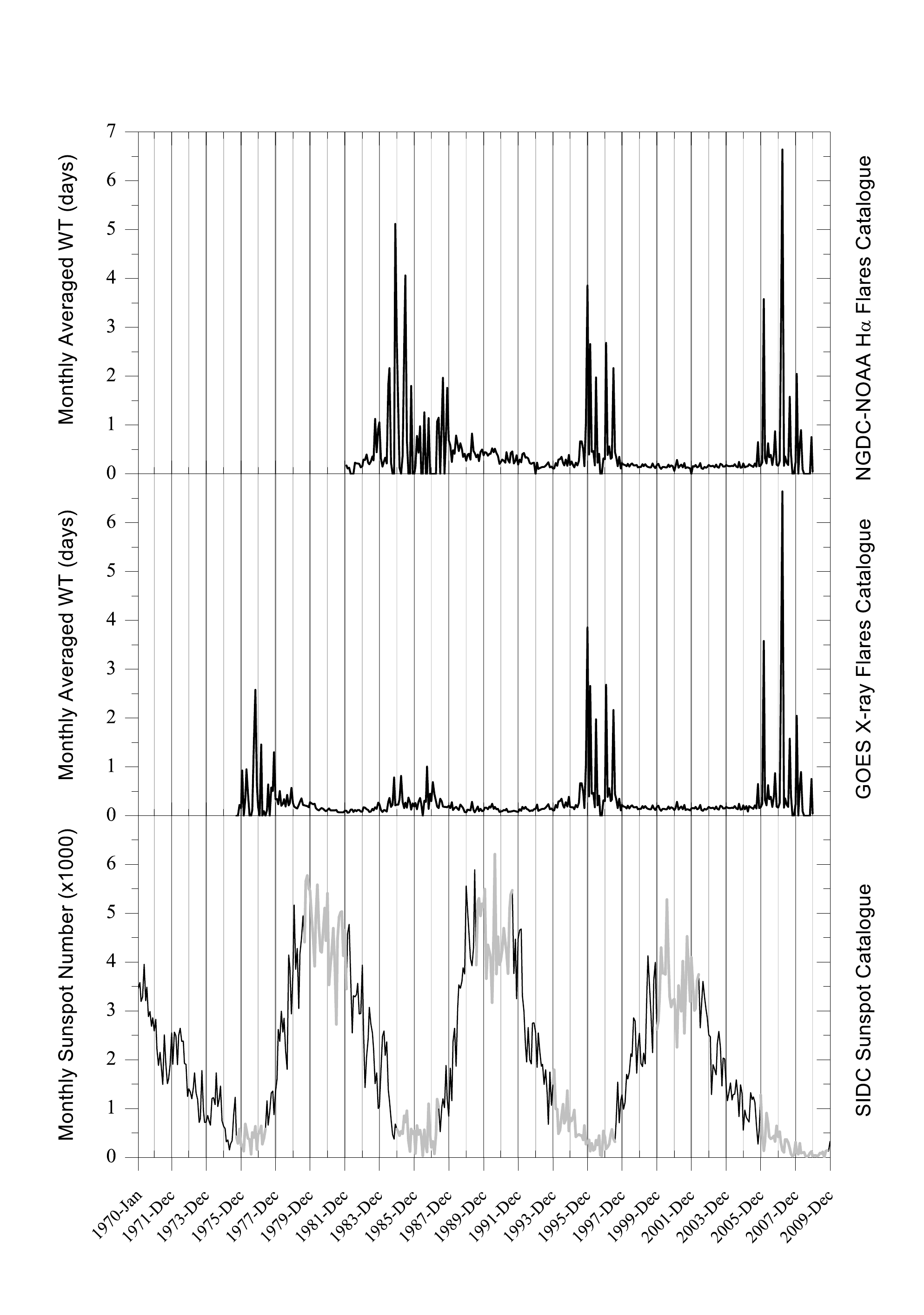}
\caption{Estimated waiting times versus monthly sunspot number for H$\alpha$ flares (top) and X-ray flares (middle). The sunspot data are provided by the Solar Influences Data Analysis Center. The solar maxima (minima) phases that have been used in the analysis are designated by thick grey lines (bottom).}\label{fig.1}
\end{figure}

The calculated \ptp waiting times from the GOES catalogue are shown in \fref{fig.1}. The correlations between monthly averaged sunspots number and monthly averaged \wt in both both Soft X-rays and H$\alpha$ are evident and intuitive. During the solar minimum years the flare rate is lower, leading to longer \wt prevalence; conversely, during the solar maximum years the rate is higher and thus waiting times are shorter on average.

This substantial variation leads to separate consideration of the solar cycle phases according to corresponding traces in the calculated \wt series. To separate phases, we consider the time derivative of the monthly average sunspot number $\dot{N}= d N(t)/dt$. The qualitative agreement (in \fref{fig.1} is shown by grey lines) in the fluctuations of \wt and $N(t)$ is attained empirically, by considering intervals with limited fluctuations in $\dot{N}$ around the corresponding zeros, to be less than $25\%$ for the minima phases and less than $50\%$ for the maxima.

\begin{figure}
\includegraphics[width=1\textwidth,clip=]{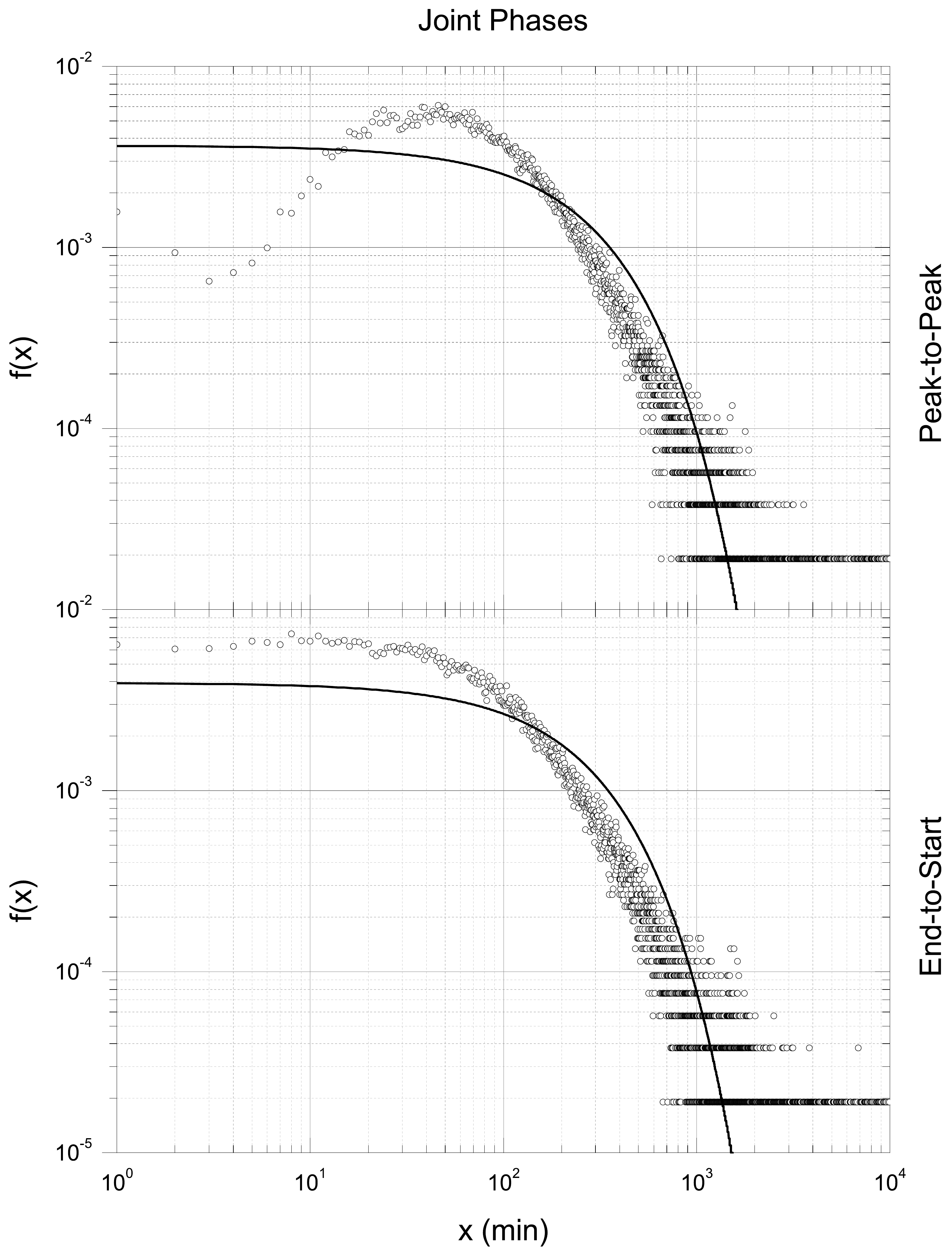}
\caption{The \pdf of the flare waiting times. The waiting times are estimated from the whole GOES Soft \Xr dataset by means of the different \wt definitions, which are shown in separate panels. The solid line is the pure exponential reference curve with the rate estimated from the data being shown by the scatter plot.}\label{fig.2}
\end{figure}

\begin{figure}
\includegraphics[width=1\textwidth,clip=]{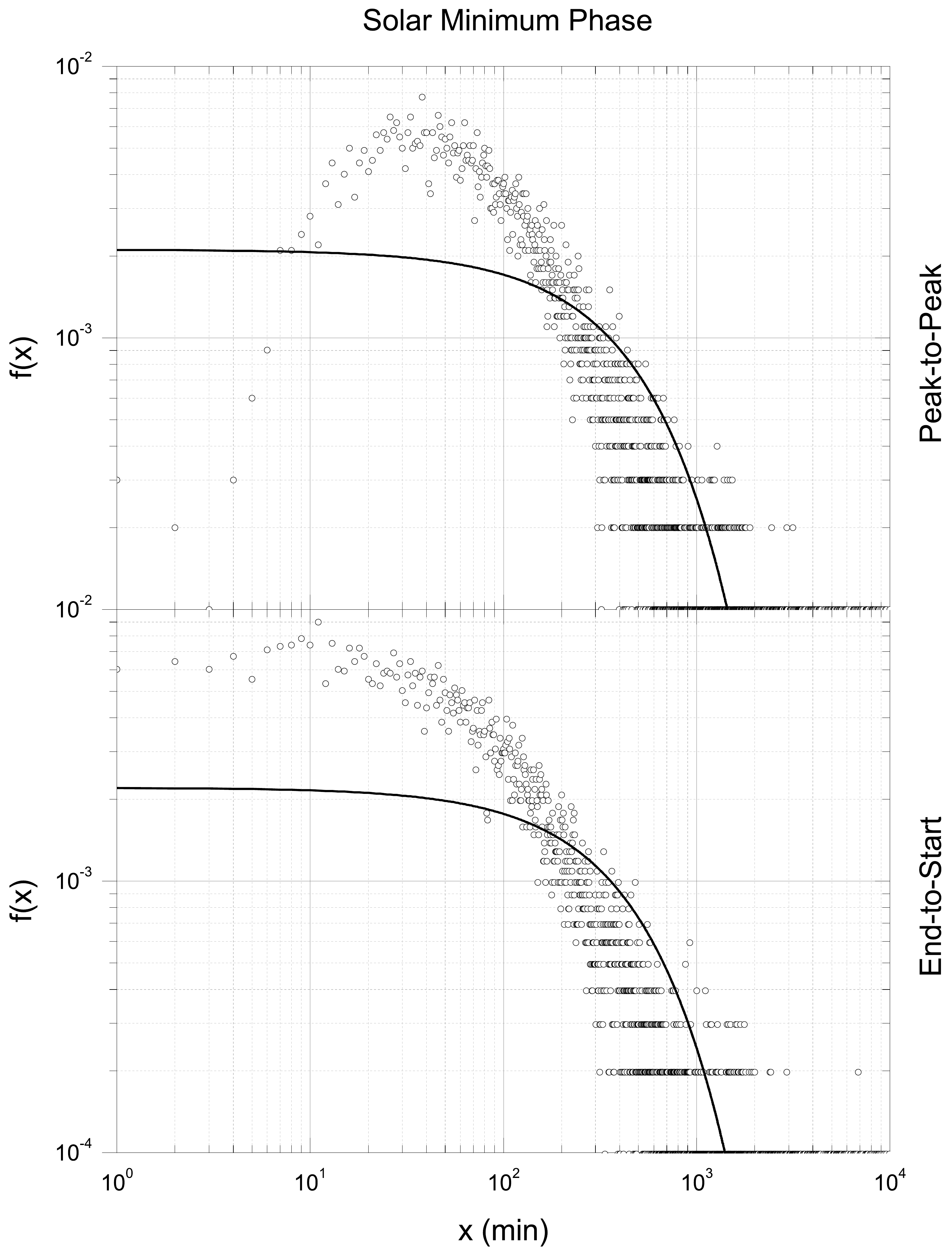}
\caption{The \pdf of the flare waiting times. The waiting times are estimated from the GOES Soft \Xr dataset reduced to the years of the solar minimum by means of the different \wt definitions, which are shown in separate panels. The solid line is the pure exponential reference curve with the rate estimated from the data being shown by the scatter plot.
}\label{fig.3}
\end{figure}

\begin{figure}
\includegraphics[width=1\textwidth,clip=]{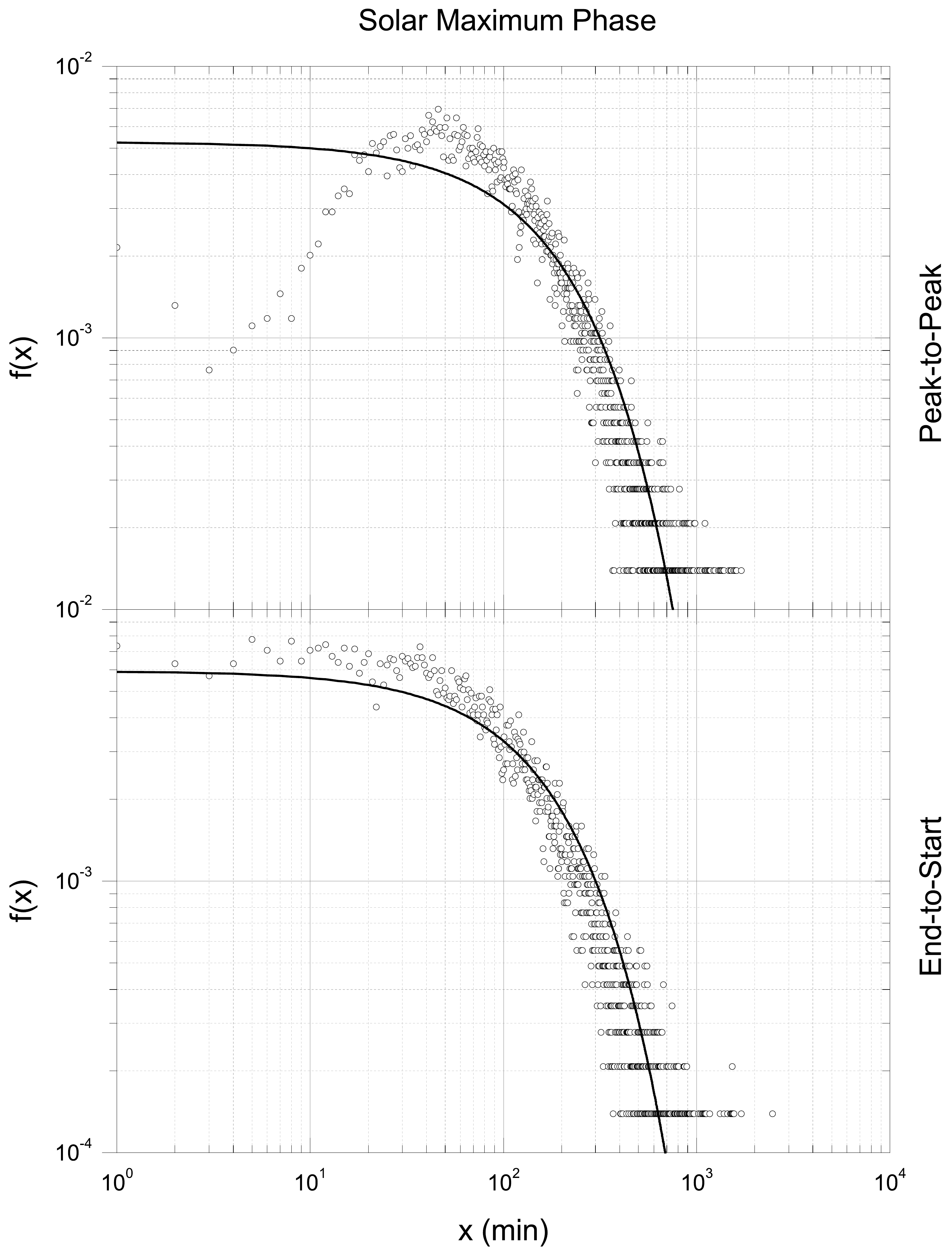}
\caption{The \pdf of the flare waiting times. The waiting times are estimated from the GOES Soft \Xr dataset reduced to the years of the solar maximum by means of the different \wt definitions, which are shown in separate panels. The solid line is the pure exponential reference curve with the rate estimated from the data being shown by the scatter plot.}\label{fig.4}
\end{figure}

Figures~\ref{fig.2}-\ref{fig.4} show a \pdf of waiting times for joint solar cycle phases, solar minimum and maximum, respectively, by means of the two \wt definitions mentioned above. The solid line in each panel stands for the reference exponential distribution $\lambda e^{-\lambda x}$, with the average $\lambda^{-1}$ estimated from the series whose \pdf is shown by the scatter plot. To some extent, the similarity with the pure exponential is attained in a single case only (the bottom panel in \fref{fig.4}).

The density shapes are similar within a given \wt definition. Remarkably, the variation in \wt definition affects the most probable events: the bell-like \pdf versus plateau-like over almost $1.5$ decades.

The plots in \fref{fig.2} should be interpreted as power-laws according to \inlinecite{Boffetta}. However, the power-law fit has to be considered with special care, as the findings illustrated in the following are challenging the straightforward applicability of the power-law fit.

\subsection{Flaring rate estimation}

\inlinecite{Wheat-II} had ignored flares whose peak \Xr flux is less than $10^{-6}$ Wm$^{-2}$, due to substantial variation of the soft \Xr background with the solar cycles. Thus GOES $B$-class events ($B$-flares for short) had not been considered.

Following \inlinecite{Wheat-II}, additional sets of data without $B$-flares were generated. In the following, the effect of this removal is considered systematically.

\begin{figure}
\includegraphics[width=1\textwidth,clip=]{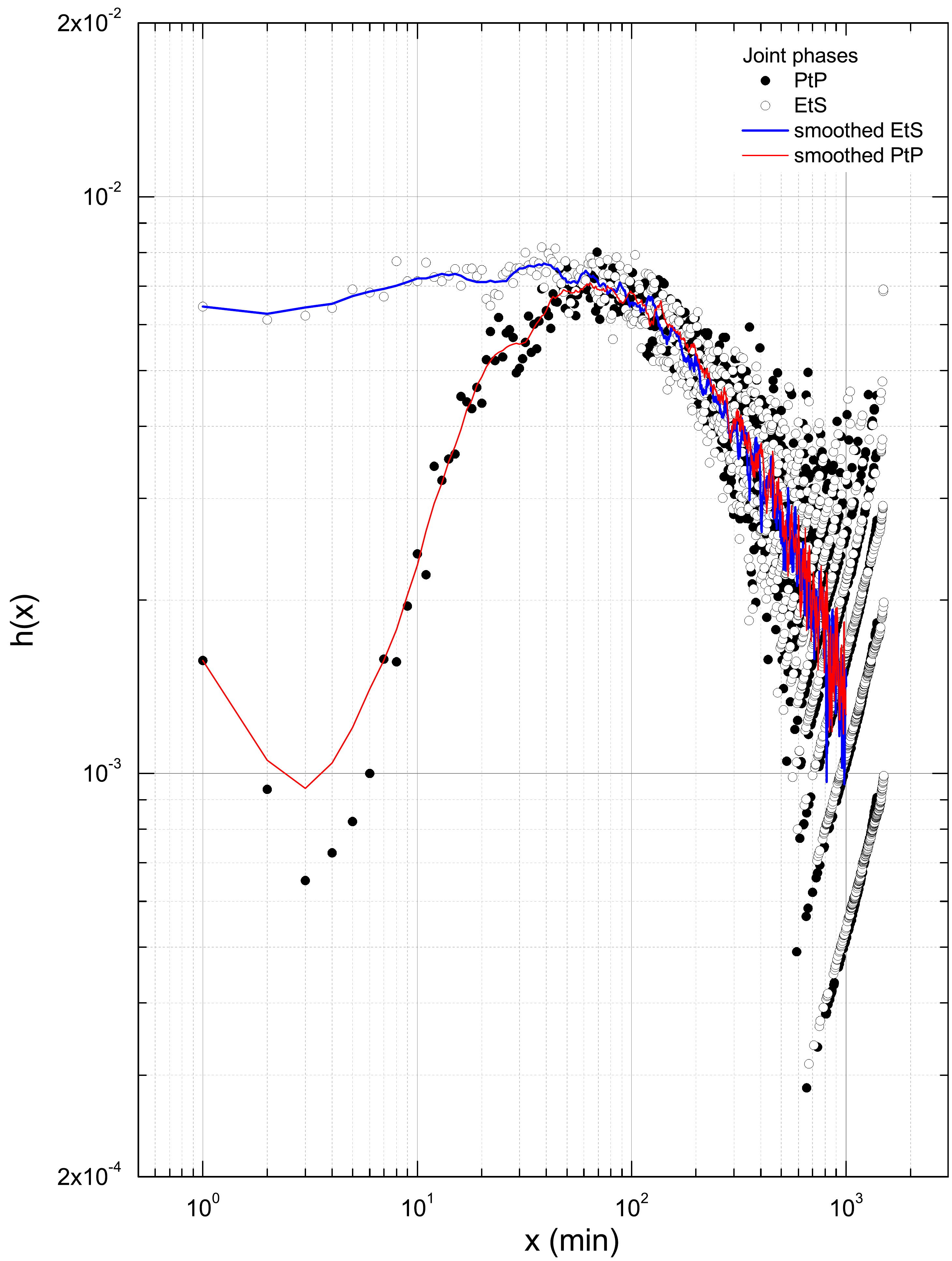}
\caption{The estimated flare rate $h$. The whole set of the GOES Soft \Xr events is used. The variation due to \wt definition is shown by the scatter plots, which differ from each other in the region of short \wt. The solid lines are the smoothing applied to emphasise the qualitative behaviour. The $B$-class flares are included.}\label{fig.5}
\end{figure}

\begin{figure}
\includegraphics[width=1\textwidth,clip=]{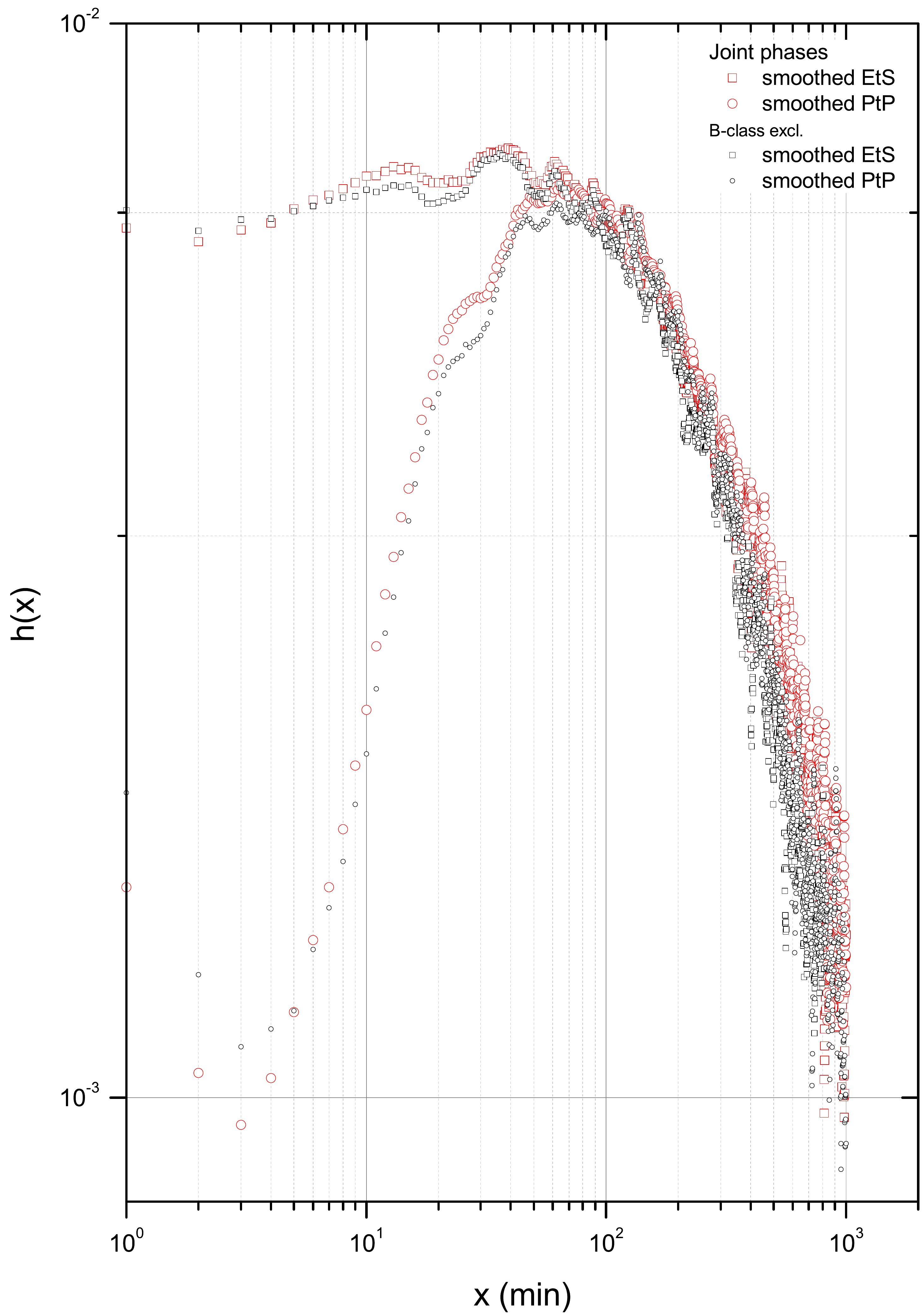}
\caption{The scatter plot of the smoothed flare rate $h$. The whole set of the GOES Soft \Xr events is used. The cases without $B$-class flares  are shown for comparison.}\label{fig.55}
\end{figure}

\begin{figure}
\includegraphics[width=1\textwidth,clip=]{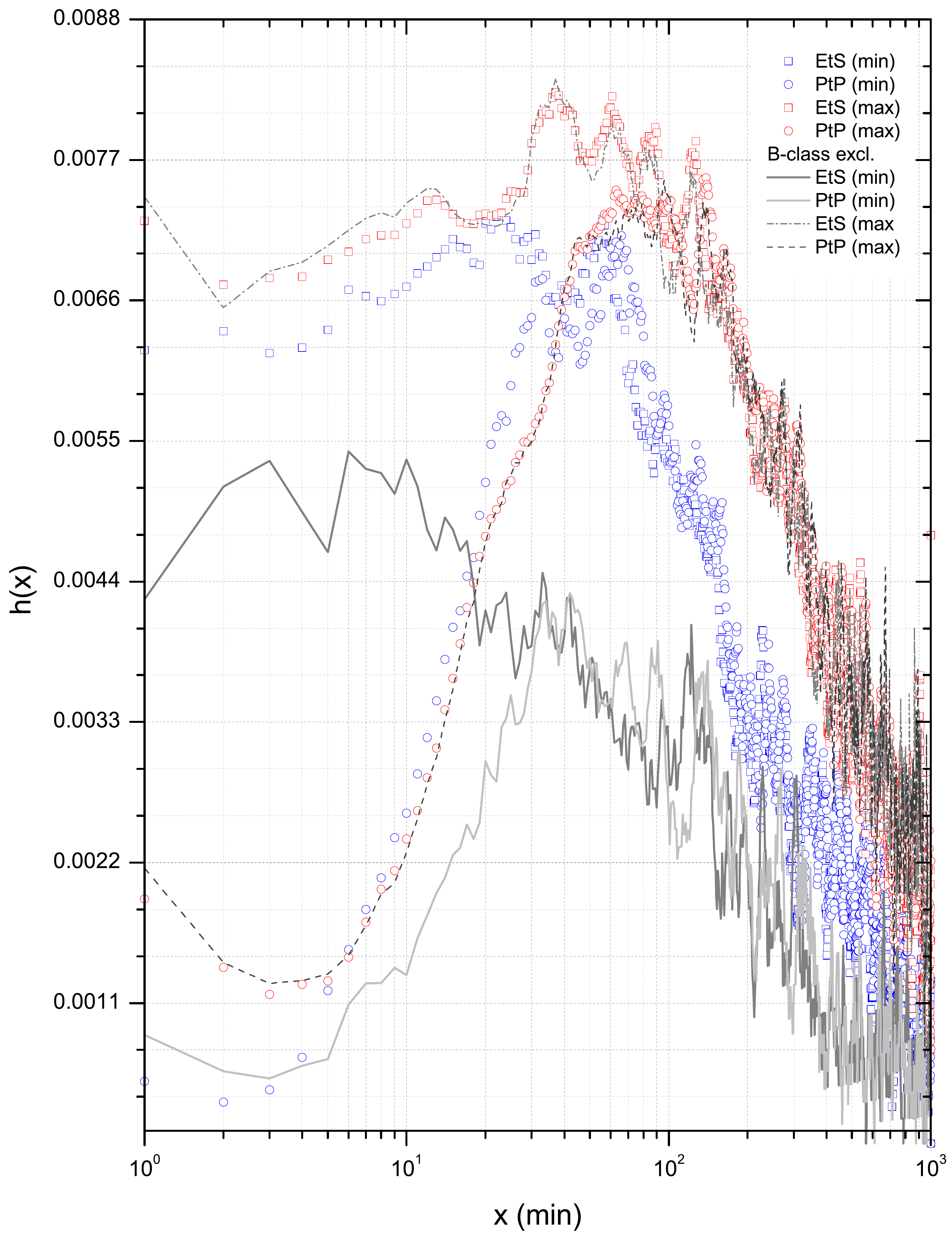}
\caption{The smoothed flare rate $h$ for the different solar cycle phases are shown separately. The cases with $B$-class flares excluded are shown by grey lines. See text for discussion.}\label{fig.6}
\end{figure}

We estimate the flaring rate $h$ explicitly from the \pdf according to \eref{main_eq}. The result for joint solar cycle phases is reported in \fref{fig.5}; in \fref{fig.55} joint phases with $B$-flares excluded are shown for comparison, and the phase-wise estimated rates are shown in \fref{fig.6}. The effect of the \wt definition on the curvature of the estimated function $h$ is shown for the entire dataset.

In \fref{fig.5}, the solid lines represent the smoothing adjacent running average with $10$-point window, to emphasise the character of the functions.

\fref{fig.55} is complementary to the previous one, and demonstrates the alteration of the smoothed rates, when the $B$-flares are removed. The smoothed rates are shown by the scatter plot.

In \fref{fig.6}, the estimated rates are shown for the different solar cycle phases separately. The estimates are smoothed by the $15$-point adjacent running average. The phases are coded by the colour of the scatter plot; the \wt definition by the symbol shapes; the cases without $B$-flares are shown by lines.

The estimated flare rate allows one to reconsider the applicability of the power-law fit for the waiting time \pdf by trivial algebra: the simplest functional form for the power-law fit is given by
\begin{equation}\label{pw}g(x)=Ax^{-\alpha},\end{equation}
\noindent
where $A$ and $\alpha$ are the estimated parameters, and due to \eref{main_eq} the corresponding failure rate to the power-law \pdf $g(x)$ reads

\begin{equation}\label{ppdf}h_g(x)= \frac{Ax^{-\alpha}}{A\int_{x}^{\infty} u^{-\alpha}du}=\frac{x^{-\alpha}}{ (\alpha-1)x^{-\alpha+1}}\approx x^{-1},
\end{equation}
with $|\alpha|>1$.

To verify this relation for the estimated rates, we fit the smoothed\footnote{The smoothing is the same as for the graphs in \fref{fig.6}.} rate $h(x)$ by a function of the form of \eref{pw}; the estimated exponent $\gamma$ is compared with $-1$.

Tables \ref{Table-ptp} and~\ref{Table-ets} report the average \wt and the exponent $\gamma$ for all datasets that had been analysed in this work.

\begin{table}
\caption{Numerical characteristics of the datasets generated according to the \ptp flare waiting time definition.}
\label{Table-ptp}
\begin{tabular}{llllll}

Catalogue&Phase &$<x>^{a}$$[h]$ & $<x>^{b}$$[h]$&\multicolumn{1}{c}{$\gamma^{a}$}&\multicolumn{1}{c}{$\gamma^{b}$}\\
\hline
   &\textit{joint} & $4.48$ & $5.89$ & $-0.86\pm0.006 $ & $-0.91\pm0.005^{c}$ \\
\Xr &\textit{max}  & $3.14$ & $3.27$ & $-0.65\pm0.008 $ & $-0.73\pm0.007^{c}$ \\
 &\textit{min}  & $9.58$ & $28.99$ & $-0.61\pm0.01 $  & $-0.67\pm0.01$    \\
\hline
   &\textit{joint} & $6.00$ & $9.87$ & $-0.54\pm0.006$ & $-0.53\pm0.008$  \\
\Ha &\textit{max}  & $4.58$ & $7.01$ & $-0.46\pm0.01$ & $-0.43\pm0.01$   \\
   &\textit{min}  & $11.15$ & $50.14$ & -$^{d}$  & --$^{d}$         \\

\hline
\end{tabular}
\tiny
\begin{enumerate}
\item[${}^a$] events including GOES $B$-class flares.
\item[${}^b$] events excluding GOES $B$-class flares.
\item[${}^c$] the power-law region is remarkably present.
\item[${}^d$] insufficient record length to produce a reliable result.
\end{enumerate}
\end{table}

\begin{table}
\caption{Numerical characteristics of the datasets generated according to the \ets flare waiting time definition. Note: the power-law regions in these data appear to be more pronounced and longer.}
\label{Table-ets}
\begin{tabular}{llllll}
Catalogue&Phase &$<x>^{a}$$[h]$ & $<x>^{b}$$[h]$&\multicolumn{1}{c}{$\gamma^{a}$}&\multicolumn{1}{c}{$\gamma^{b}$}\\
\hline
    &\textit{joint}  & $4.15$ & $5.54$  & $-0.82\pm0.005$ & $-0.99\pm0.006^{c,d}$ \\
\Xr &\textit{max}   & $2.79$ & $2.92$  & $-0.65\pm0.01$ & $-0.78\pm0.009^{c}$  \\
    &\textit{min}   & $9.33$ & $28.72$ & $-0.66\pm0.01$ & $-0.56\pm0.007$    \\
\hline
    &\textit{joint}  & $5.68$ & $9.54$  & $-0.50\pm0.005$ & $-0.52\pm0.006$ \\
\Ha\ &\textit{max}   & $4.20$ & $6.68$  & $-0.46\pm0.01$ &$-0.45\pm0.01$  \\
    &\textit{min}   & $10.94$ & $49.75$ & -$^{e}$     & -$^{e}$       \\
\hline

\end{tabular}
\tiny
\begin{enumerate}
\item[${}^a$] events including GOES $B$-class flares.
\item[${}^b$] events excluding GOES $B$-class flares.
\item[${}^c$] the power-law region is remarkably present.
\item[${}^d$] variations of the fit parameters may lead to the $-1$ power-law index.
\item[${}^e$] insufficient record length to produce a reliable result.
\end{enumerate}
\end{table}

\section{DISCUSSION}\label{discussion}

In this section, the opposing results and interpretations existing in the literature are compared and discussed in the context of the findings reported. We shall highlight the discrepancies and agreement with those previously published by the cited authors.

\subsection{Uncertainties}

Evidently, our results are strongly affected by the completeness and accuracy of the GOES catalogue. In fact, the event detection is done in an automated way and any change in the detection algorithm would alter the statistics. Furthermore, the flare event \textit{obscuration} (for details see \opencite{Wheat-Rates-Ind-Reg}) brings about quite hard arguments against the reliability of any study: the missing of substantial amount of events (up to $75\pm23\%$ for events above GOES $C1$-class) might be so crucial that the \pdf would dramatically change its behaviour. However, our results are comparable with those based on the same datasets.

In the case of the solar minima, the longer record is required most of all. The relatively high probability of longer \wt (\fref{fig.3}) conceals the probability variation along the domain, since the \pdf is shrunk along the ordinate and has too wide spreading of the tail. The longer series should clarify the \pdf shape.

\eref{main_eq} diverges as the cumulative distribution approaches values at the very tail
\begin{equation}\label{lim}\lim_{x\rightarrow\infty}F(x)=1\:,\end{equation}
and, consequently, the wide tails of the \pdf are amplified by means of the non-linear transformation given by \eref{main_eq}. Thus, the resulting tails of the estimated rates appear to be even more scattered (\fref{fig.5}). This motivated us to apply a smoothing to the functions obtained, which, in turn, eliminates details and provides rather qualitative result.

\subsection{Waiting time definition}
A random event\footnote{Here by "event" we mean the event as defined in Probability Theory.} should be defined as a point-like instance in time \textit{i.e.}, the event has no length. From this point of view, the \ptp definition is more mathematically adequate, rather than the \ets waiting time defined by the subtraction of the flare duration from the base time line. Practically, the flux maximum point is easier to detect precisely, with respect to the starting/ending times, since their flux values are closer to the background noise level.

\inlinecite{Wheat-I} and \inlinecite{Boffetta} used the \wt defined as the difference between the times of peak flux rates of two near-in-time flares. Later in \inlinecite{Wheat-II} the \wt was defined as the difference between the start times. However, the latter definition is qualitatively coincident with what we called \ptp. We would ignore possible numerical discrepancies.

The major differences in the estimated rates caused by the \wt definition had appeared in the range of short waiting times (Figures~$5-7$).

In spite of the cut of wide-spreading tails, according to the authors the convergence of rates at longer waiting times can be identified, regardless of the \wt definitions that had been used. This is seen where the tails of the flare rates seem to coincide. Such a convergent behaviour is independent of the solar cycle phase, as well as of the $B$-flares exclusion (\fref{fig.6}). However, the $B$-flares exclusion leads to more ambiguous plots, since fewer events are considered in the dataset.
The joint consideration of the phases demonstrates this behaviour too~(\fref{fig.5}).

If we assume that energetically large flares are separated by longer waiting times on the average, we can conclude that the relatively long duration of energetic flares appears to be statistically indistinguishable from the long waiting times defined by the point-like events. Thus, the \ets definition tends to mimic the \ptp in the right-hand half of the $x$ domain, where the waiting times are longer.

\subsection{Waiting time pdf versus flaring rate}

In this work, the \pdf are presented mostly for illustrative purposes, \textit{i.e.} to demonstrate a degree of divergence between the exponentials, the variation due to the solar cycle phase and the effect of the \wt definition. We do not consider the \pdf fit, but study the underlying models by means of the estimated flaring rate.

The failure rate formalism described in \sref{math} leads to a broader view on the solar flaring and physics underlying the waiting time statistics. The function $h(x)$ permits a qualitative description of the dynamics of the stochastic process being characterised by the pdf $f(x)$. In fact, the rate $h(x)$ and the pdf $f(x)$ are related to different stochastic processes: the rate $h(x)$ is the almost instantaneous probability of an event\footnote{per unit time ($minute$).} with no event during the waiting time $x$, by the definition. In turn, the \pdf $f(x)$ is the probability change rate of the waiting time $x$. The function $h(x)$, by definition, explicitly combines the almost instantaneous probability of the flare with the waiting time that had elapsed before its occurrence, \textit{i.e.} it is the conditional probability.

In other words, the rate $h(x)$ gives the probability of the event delimiting the length of $x$, namely it is related to the process \textit{inducing} the one being presented by the series of the waiting times in the catalogue.

In the present work, we use the estimated rates $h$ to analyse the flaring dynamics which can be derived from the GOES catalogue, and then we reconsider the cited stochastic models of the waiting time pdf fit. Particularly, in the piece-wise constant Poisson model the flare rate is a free parameter.

\subsection{Non-constancy over the waiting time domain}

The most important feature of the estimated $h$ is the variation over the $x$ domain. The function $h$ is the explicit non-linear function of $x$:\begin{equation}\label{var}h=h(x)\end{equation}
with bell-like shape in semi-logarithmical plot. This has far-reaching consequences for the flaring dynamics.

First, the flaring rate is by no means stationary: it non-linearly depends on waiting time of short scales of days, hours and minutes (the fast non-stationarity for short).

Second, the flare rate has a certain behaviour: less energetic flares ($B$-flares at large, but not only these) exhibit an \textit{increasing} rate. Next, the rate reaches \textit{maximum} values, and then it rather slowly \textit{declines} (similarly to a power-law, which in some cases manifests itself very notably). Thus, we can point out \textit{characteristic waiting times}, which indicate the time intervals of the most probable flare occurrence. We would not define this time as numerically accurate due to significant uncertainties. However, one can empirically define a range just by examining \fref{fig.55} and \fref{fig.6}: for joint phases, say~35--45~$min$. (\ets); 50--60~$min.$ (\ptp); for the solar minimum, say 20--55~$min$. (\ets); 50--75~$min$. (\ptp); and for the solar maximum 30--45$~min.$ (\ets); 60--100~$min.$ (\ptp).

The character of the fast non-stationarity is invariable within the solar cycle phases (\fref{fig.6}), and remains recognisable for smaller data records, when the $B$-flares are excluded (with a word of caution on the reliability of the result based on  the smaller number of samples).

\subsection{Rate change with the solar cycle phase}

The slow non-stationarity of the flare rate is given by the solar cycles, say "large scale". The phases have different probability of flaring per unit time, as it is natural to assume just on the basis of the comparative overview in \fref{fig.1}.

The most frequent and thus very short flares (corresponding to the smaller $x$) are common features for both phases. They can hide statistical differences between solar cycle phases: this fact is supported by the match in the increasing regions of the rates corresponding to the disjoint phases. In \fref{fig.6}, the \ptp rates are equivalent in the range from $4$ to $20~min.$, regardless of the solar cycle phase considered. The same statement but somewhat weaker holds for the \ets rates.

Qualitatively, the rate shape is invariable during the solar cycle, however, the variation in the mean \wt may reach one order of magnitude. This is the case when the $B$-flares are excluded (see Tables~\ref{Table-ptp} and \ref{Table-ets}).

Summarising, the flare rate takes the form \begin{equation}\label{hxt}{h=h(x,t)}\end{equation} with explicit dependence on waiting time and generic time, \emph{i.e.} the time frame when the corresponding observations were made.

\subsection{GOES $B$-class events}

We considered separately the case of the excluded $B$-flares. For joint phases almost $32\%$ are $B$-flares and $3.4\%$ for the solar maxima.

Qualitatively, the rates retain their shape over the \wt domain. Nevertheless, the rates without $B$-flares are systematically lower in the joint phases (\fref{fig.55}) and with almost negligible variations during the solar maxima (in \fref{fig.6} dashed and dash-dotted lines).

Substantial variation occurs during the cycle minima phases, when $80\%$ of the events are $B$-flares. In this case, the notable rate matching in the shorter increasing range is very small for the \ptp rate, and does not takes place for the \ets rates (in \fref{fig.6} solid grey lines).

In fact, the elimination of the frequent events from the dataset is a very significant operation. Particularly, the power-law exponents are sensitive to the relative strength of the frequent events. This is caused by the intrinsic divergence of the power-law statistics at the origin, at the most probable (\emph{i.e.} the most frequent) events.

In some cases, excluding $B$-flares decreases the reliability of the results because of the small size of the resulting dataset. Nevertheless, it underlines the average waiting time order of magnitude change, which is the reason for the separate considerations of the solar cycle phases (see Tables~\ref{Table-ptp} and \ref{Table-ets} for numerical estimates).

\subsection{Power-law fit}
Considering \eref{ppdf}, we compare the flare rate power-law fit with the value of $-1$.

One should notice that the tails of $h(x)$ are quite uncertain in considering the indices listed in Tables~\ref{Table-ptp} and \ref{Table-ets} to be steady and exact: the change of the region boundaries being chosen for the fit can substantially modify the numerical value of the exponent, and, in a certain sense, it may be considered to be subjective.

But we rely on the fact that the modulus of the estimated indices is systematically \textit{less} than $1$. Thus, arguments for fitting the GOES data by \eref{pw} can, very likely, be rejected.

On the other hand, jointly considered solar cycle phases in the \Xr band with flares above $B$-class have revealed a value comparable with $-1$. By noting, that this takes place when average waiting times differ by one order of magnitude, we would point the accordance with the power-law limit \pdf of the non-stationary exponential random variable, reported by \inlinecite{Wheat-final}. This argument is missing in \inlinecite{Boffetta}, when joint phases have been fitted confidently by the power-law.

\subsection{Excess of short events}

The excess of short waiting times was pointed out by \inlinecite{Pearce} and \inlinecite{Wheat-I}. It appears to be an intrinsic property of the flaring process, and it is detectable regardless of the instrument and(or) the band that is used.

The rising character of the rate $h$ at small $x$ appears to be an indicator of the importance of the energetically small events. Short waiting times mostly separate less energetic events. Empirically, this point is supported by the influence of the \textit{B}-flares' presence on the flare rate. In turn, the increasing probability at \textit{short} \wt suggests that the true flare occurrence per short time unit is very large, but it is hidden due to sensitivity of the instruments and the obscuration. Certainly, somewhere on these time scales the very sympathetic flaring \cite{Moon,Biesecker} would take place, and one can realise that a dependence of the form $h=h(x)$ poses explicitly the "memory" in the record\footnote{Recall that memoryless stochastic processes are those with $h=const$, \textit{i.e.} the probability of an event is independent of neither the waiting time elapsed before it occurs nor the preceding event.}. Thus, it is hard to ignore the arguments by \inlinecite{Lepreti} concerning the memory effect in the waiting time pdf. That is also in accordance with the divergence (\emph{viz.} short flare overabundance) of the flare duration \pdf at the origin, which is commonly accepted to be a power-law.

The GOES soft \Xr flare catalogue gives an overall flaring picture, considering the Sun as a whole. The beauty and power of the non-stationary Poisson model is in its potential ability to represent the global solar flare dynamics as consisting of somewhat trivial ("memoryless") mathematical objects --- exponentials. Presumably, the different active regions (or even smaller regions) may contribute to the global flaring with notably different rates and thus appear in the limit as a power-law-like distribution. However, the individual active regions have very poor statistics (about 100 events at most) and a series of assumptions should be made \emph{a priori} (\opencite{Wheat-Rates-Ind-Reg}). In addition, \inlinecite{Wheat-Rates-Ind-Reg} reported a piece-wise constant Poisson fit for the active region, which gives a motivation to consider even smaller flaring areas to be an "elementary piece" in the mechanism just speculated on above.

\section{CONCLUSIONS}\label{conclusion}
The fast variation on time scales from minutes to hours of the flare rate is a highlight among other findings in this work. The large-scale variation with the solar cycle exhibits a complexity that can be coped with by a phase-wise splitting of the data. However, the fast non-stationarity during the waiting time appears to be an intrinsic feature of the flaring dynamics, which requires a further elaboration of the present solar flaring models.

It is worth mentioning that we see somewhat intermediate character of the reported results with respect to the works by \inlinecite{Wheat-final}, \inlinecite{Boffetta} and by \inlinecite{Lepreti}. The "memory" revealed in the data rejects models with Poisson statistics. From another perspective, a simple power-law fit very likely fails for the records relevant to a specific solar cycle phase.

The piecewise-constant Poisson model had inspired us to consider solar cycle phases separately. We support the argument of this model that solar cycle phases having quite inhomogeneous statistical properties that are to be considered jointly. In other words, the separate consideration of the solar cycles should be a cornerstone of a realistic modelling of the solar flaring activity.

On the other hand, our findings exclude the presence of the timescale of true rate constancy (with GOES catalog precision, at least). Even if it exists, it would be very small, and may be hard to detect.

The application of Renewal Theory emphasises the importance of the short (energetically small) events in the pdf, whose dynamics had not been pointed out in the literature: partially due to removal from the dataset of the $B$-class flares, and partially due to systematic fitting solely the tails of the \pdf's. This brings us to problems similar to those that had arisen in the context of coronal heating by nanoflares, where, perhaps, the most significant effects are at the limit of the instrumental noise.

\begin{acks}
   We thank the GOES teams at NOAA and SIDC for data management and availability and Christoph Keller for useful comments and discussions. M.M. acknowledges the support of the Italian Space Agency (ASI) and COST Action ES0803. Mrs. S. Fabrizio (INAF-OATS) is gratefully acknowledged for careful proofreading.
\end{acks}

\clearpage

\end{article}

\end{document}